%% file: paper_full.tex
\documentclass{llncs}
\usepackage{todonotes}
\usepackage{wrapfig}
\usepackage{hyperref}
\usepackage{xspace}

\newcommand{\pleak}{\textsc{Pleak}\xspace}

\title{Business Process Privacy Analysis in \pleak}

\author{Aivo Toots\inst{1,2} \and Reedik Tuuling\inst{1} \and Maksym Yerokhin\inst{2} \and Marlon Dumas\inst{2} \and Luciano Garc{\'i}a-Ba{\~n}uelos\inst{2} \and Peeter Laud\inst{1} \and Raimundas Matulevi\v{c}ius\inst{2} \and  Alisa Pankova\inst{1} \and Martin Pettai\inst{1} \and Pille Pullonen\inst{1,2} \and Jake Tom\inst{2}}
\institute{
Cybernetica AS, Estonia\\
\email{firstname.lastname@cyber.ee}\and
University of Tartu, Estonia\\
\email{firstname.lastname@ut.ee}
}

\begin{document}

\maketitle

\begin{abstract}
\pleak{} is a tool to capture and analyze privacy-enhanced business process models to characterize and quantify to what extent the outputs of a process leak information about its inputs. \pleak{} incorporates an extensible set of analysis plugins, which enable users to inspect potential leakages at multiple levels of detail. 
\end{abstract}

\section{Introduction}
\input{intro}

\section{PE-BPMN Editor and Simple Disclosure Analysis}
\input{inn-simple-disc}

\section{Qualitative Leaks-When Analysis}
\input{inn-leaks-when}

\section{Sensitivity Analysis and Differential Privacy}

\input{inn-sensitivity-full}

\label{sec:sensitivity}

\section{Attacker's Guessing Advantage}

\input{inn-advantage-full}
\label{sec:advantage}

\bibliographystyle{abbrv}
\bibliography{paper}

\appendix
\newpage
\section{Demo Plan}
\input{demo-plan}

\end{document}

%% file: intro.tex
% !TEX root = ../paper.tex

% Numerous recent incidents involving the disclosure of personal data have
% heightened society’s awareness of the vulnerability of private information 
% within the cyberspace. Initiatives such as the
% General Data Protection Regulation (GDPR) by the European Union and
% the new Privacy Shield agreement in the United states have seen light,
% with the aim of strengthening data security and preventing privacy breaches.

% In this context, the NAPLES project aims at developing fundamental research and
% methods to enable the analysis of potential data leakages over business processes.
% To that end, an extension to the BPMN standard notation has been proposed to
% allow the specification of privacy-enhancing technologies overlaid over BPMN
% process diagrams. 

Data minimization is a core tenet of the European General Data Protection Regulation (GDPR)~\cite{Colesky2016}. According to GDPR, usage of private data should be limited to the purpose for which it has been collected. 
To verify compliance with this principle, privacy analysts need to determine who has access to the data and what private information these data may disclose.
%, not only directly via the input data of the tasks they perform, but also indirectly via the chain of tasks that lead to the production of these data. 
%In the context of business processes, this principle implies that workers involved in the execution of a business process, should only have access to private data to the extent it is required to perform their tasks. 
Business process models are a rich source of metadata to support this analysis. Indeed, these models capture which tasks are performed by whom, what data are taken as input and output by each task, and what data are exchanged with external actors. Process models are usually captured using the Business Process Model and Notation (BPMN). 

%For example, BPMN can be used to describe how data moves through local process as well as data exchanges between stakeholders. Hence, BPMN models are a good starting point to discuss privacy issues. 

This paper introduces \textsc{Pleak}\footnote{\url{https://pleak.io} (account: \emph{demo@example.com}, password: \emph{pleakdemo}, manual: \url{https://pleak.io/wiki/}, source code: \url{https://github.com/pleak-tools/})\\This research was funded by the Air Force Research laboratory (AFRL) and Defense Advanced Research Projects Agency (DARPA) under contract FA8750-16-C-0011. The views expressed are those of the author(s) and do not reflect the official policy or position of the Department of Defense or the U.S. Government.} -- the first tool to analyze privacy-enhanced BPMN models in order to characterize and quantify to what extent the outputs of a process leak information about its inputs. 
The top level, namely the Boolean level (Sec.~\ref{sec:pe-bpmn}), tell us whether or not a given (intermediate or final) output of a process may reveal information about a given input.
%is based on cryptographic privacy enhancing technologies (PETs) and differentiates between public and protected data, where public data may cause privacy risks. 
The middle level, the qualitative level (Sec.~\ref{sec:leaks-when}), goes further by indicating which attributes of (or functions over) a given input data object are potentially leaked by each output, and under what conditions this leakage may occur.
The lower level (quantitative analysis) quantifies to what extent a given output leaks information about an input, either in terms of a sensitivity measure (Sec.~\ref{sec:sensitivity}) or in terms of the guessing advantage that an attacker gains by having the output (Sec.~\ref{sec:advantage}).

\begin{figure}[h]
    \centering
    \includegraphics[width=0.9\textwidth]{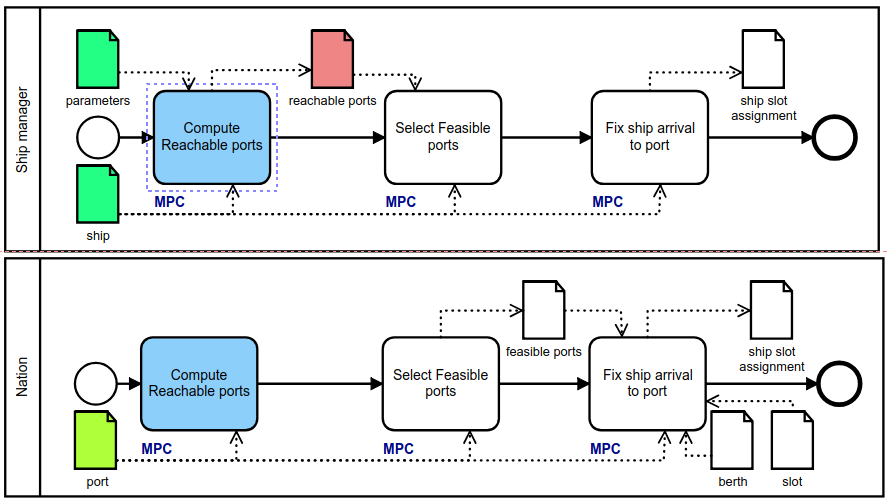}
    \caption{Aid distribution process}
%    \caption{BPMN model with MPC stereotypes highlighting the first MPC task where the nation and ship manager jointly compute ports reachable for some ship.}
    \label{fig:pe-bpmn-model}
    \vspace*{-4mm}
\end{figure}

To illustrate the capabilities of \pleak, we refer to an ``aid distribution'' process in Fig.~\ref{fig:pe-bpmn-model}. 
This process starts when a nation requests aid from the international community to handle an emergency and a country offers to route a ship to help transport people and/or goods. The goal of the process is to allocate a port and a berth to the ship but not to reveal information about ships that are unable to help or the parameters of the ports.
The process uses a type of privacy-enhancing technology (PET) known as secure multiparty computation (MPC). MPC allows participants to perform joint computations such that none of the parties gets to see the data of the other parties, but can learn the output depending on the private inputs. Given a ship, a deadline and the list of ports, task ``Compute reachable ports'' retrieves the list of ports reachable by the deadline. Tasks with identical names in different pools denote MPC computations carried out jointly by multiple stakeholders.
Task ``Select feasible ports'' retrieves ports with the capacity to host the ship. The third task selects a port, a berth, and a slot for the ship, and discloses them to both participants.  

%PE-BPMN is an extension of BPMN that allows one to attach \emph{stereotypes} to elements of a BPMN model in order to associate these elements with PETs. For example, Fig.~\ref{fig:pe-bpmn-model} includes six associations annotated with the MPC stereotype, which denotes that an input object is used by a task that is part of a collaborative computation. Other PETs supported by PE-BPMN include (homomorphic) encryption, differential privacy and secure enclave computation.

%PE-BPMN, introduced below.
%In the following, we use this scenario to illustrate some of the analyses provided by \pleak tools.

%\paragraph{Access to \pleak}

%% file: inn-simple-disc.tex
% !TEX root = ../paper.tex

\enlargethispage{0.4\baselineskip}
\label{sec:pe-bpmn} %,PTMT18
The model in Fig.~\ref{fig:pe-bpmn-model} is captured Privacy-Enhanced BPMN (PE-BPMN)~\cite{PMB2017}. PE-BPMN uses stereotypes to distinguish used PETs, e.g. MPC or homomorphic encryption, that affect which data is protected in the process. The PE-BPMN editor allows users to attach stereotypes to model elements and to enter the stereotype's parameters where applicable. The editor integrates a checker, which verifies stereotype specific restrictions. For example, that: (1) when a task has an MPC stereotype, there is at least one other ``twin'' task with the same label in another pool, since an MPC computation involves at least two parties; (2) when one of these tasks is enabled, the other twin tasks is eventually enabled; and (3) the joint computation has at least one input and one output. 
%The PE-BPMN editor reports any syntactic errors or enables privacy analysis on correct model. The error list provides visual feedback to analysts to identify the elements involved in the problem, thus helping in fixing them.

\begin{figure}[hbt]
\centering
\includegraphics[width=0.8\textwidth]{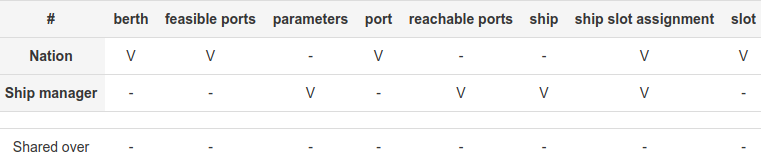}
\caption{Simple disclosure report for the aid distribution process in Fig.~\ref{fig:pe-bpmn-model}}
\label{fig:simple-disclosure}
\vspace*{-4mm}
\end{figure}

Given a valid PE-BPMN model, \pleak runs a binary privacy analysis, which produces a \emph{simple disclosure report} and data dependency matrix. The disclosure report in Fig.~\ref{fig:simple-disclosure} tells us whether or not a stakeholder gets to see a given data object. In the report shown . ``V'' indicates that a data object (in columns) is visible to a stakeholder (in rows). Row ``shared over'' refers to the network service provider, who may also see some of the data (e.g. unencrypted data objects). %The data dependency matrix shows that the \emph{reachable ports} data seen by the ship manager also depends on the \emph{port} data that is private data of the Nation. 
%For each visible data object the analyst may study it further using the followig next analyzers.

%% file: inn-leaks-when.tex
% !TEX root = ../paper.tex

\label{sec:leaks-when}
Leaks-When analysis~\cite{DumasGL2018} is a technique that takes as input a SQL workflow and 
determines, for each (output, input) pair such that the output discloses information about the input, which attributes of the input object are disclosed by the output object and under which conditions. 
A SQL workflow is a BPMN process model in which every data object corresponds to a database table, defined by a table schema, and every task is a SQL query that transforms the input tables of the task into its output tables. 
\begin{figure}[hbt]
	\centering
    \includegraphics[width=1\textwidth]{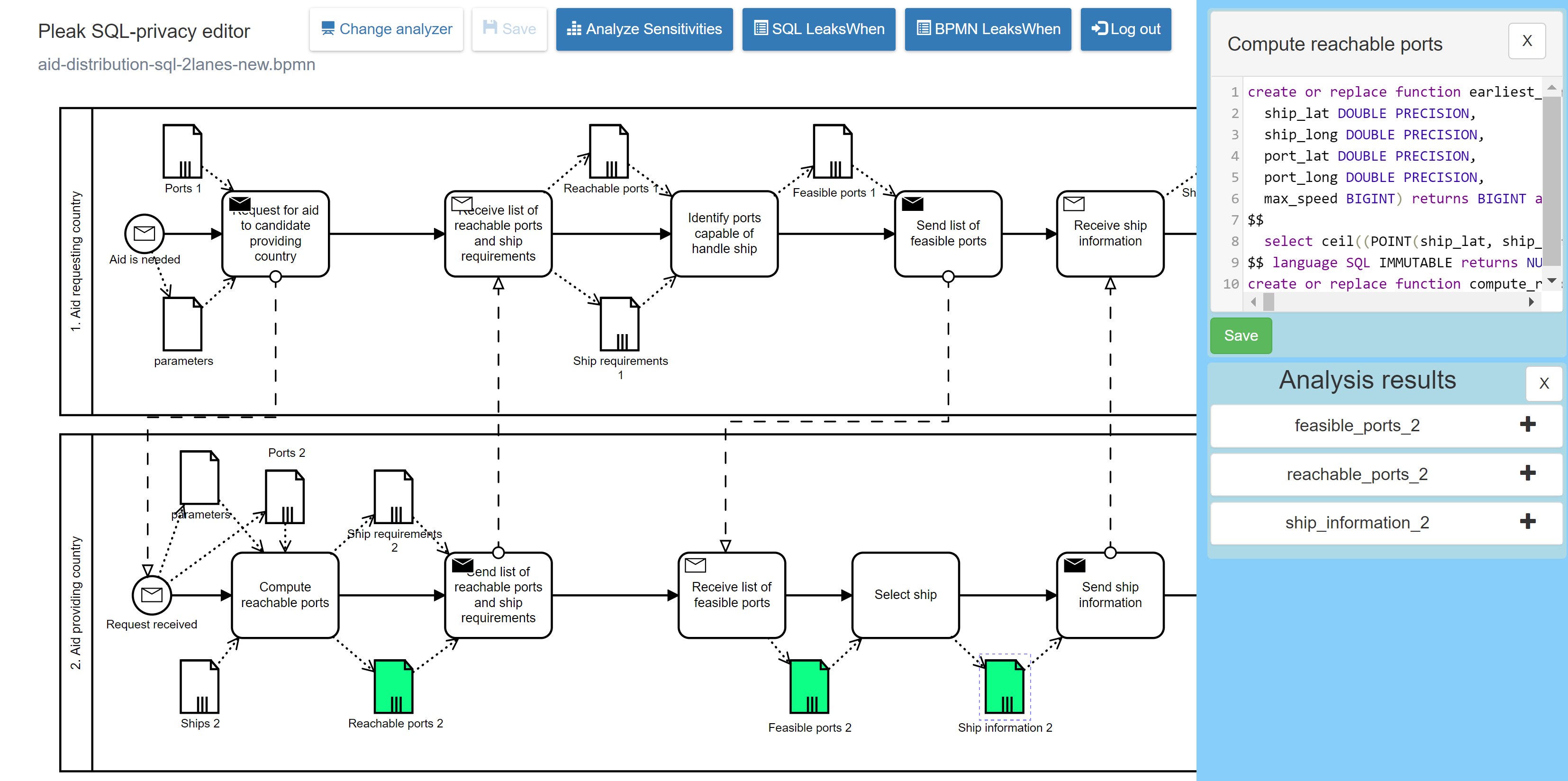}
    \caption{\label{fig:leaks-when}Aid distribution SQL workflow in \pleak SQL editor}
    \vspace*{-4mm}
\end{figure}

Fig.~\ref{fig:leaks-when} shows a sample SQL workflow -- a variant of the 
``aid distribution'' example where the disclosure of information about ships to the  aid-requesting country is made incrementally. The figure shows the SQL workflow alongside the query corresponding to task ``Select reachable ports''.

%The Leaks-When analyzer takes as input SQL collaboration workflows, as the example above, that capture information exchanges between participants. It is also possible to analyze SQL workflows comprising a single process, i.e. no participants nor information exchange  explicitly specified, such that the Leaks-When would characterize data disclosure as  side effect of computation steps and dismissing data exchanges from the analysis.

\begin{wrapfigure}{R}{0.49\textwidth}
    \vspace{-13mm}
    \begin{center}
      \includegraphics[width=0.48\textwidth,clip]{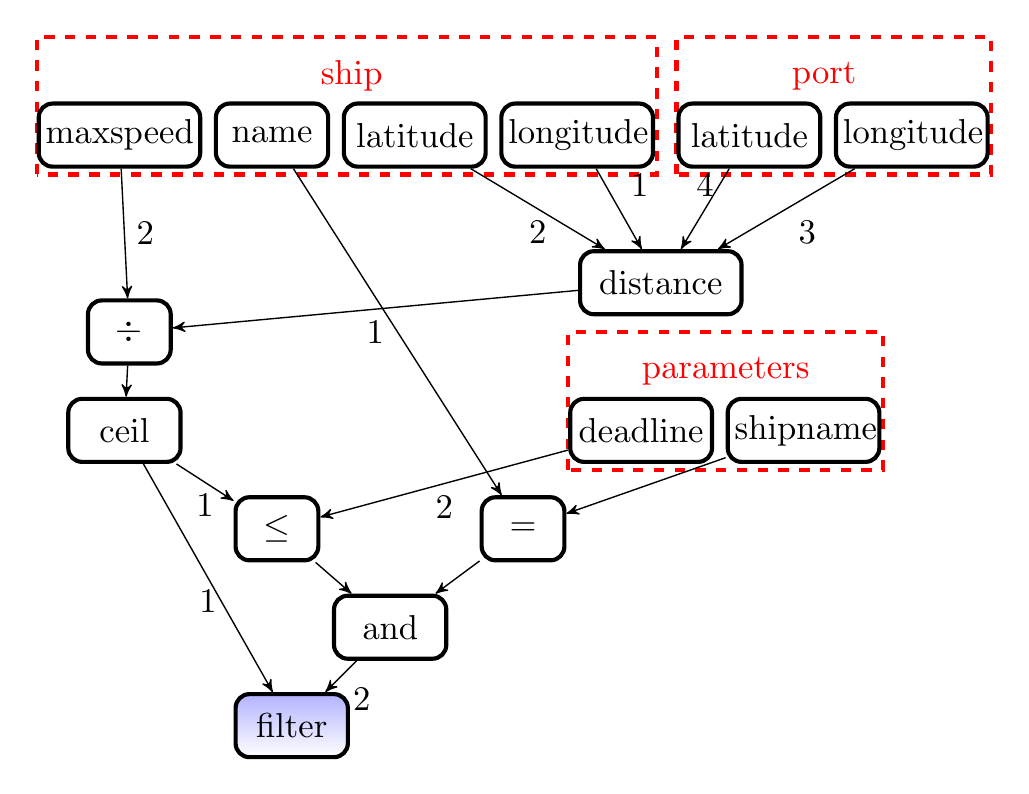}
    \end{center}
    \vspace{-5mm}
    \caption{\label{fig:leaks-when-graph}Sample leaks-when report}
    \vspace{-8mm}
\end{wrapfigure}

% process developed as part of DARPA's Brandeis project. In the example, it is assumed that a 
% country in an emergency situation requests for aid to ships (from other countries) that are 
% located close to harbor ports of the aid-requesting country. Moreover, it is assumed that the each
% country will prefer to keep undisclosed sensitive information such as the exact 
% location of the ships. 

To perform a Leaks-When analysis, the user selects one or more output data objects and clicks the ``LeaksWhen Report'' button. The Leaks-When analysis  shows one tab for each output data object and one report for each column in the output table. An example of a leaks-when report (in graphical form) is shown in Fig.~\ref{fig:leaks-when-graph}. The report states that the aid-requesting country would get to know that at least one or several ships (left branch) can reach a specific port (right branch) before the deadline (branch in the middle). The rest of the report specifies how the disclosed elements are computed from the inputs (in the dashed rectangles). The report is generated by extracting all runs of the workflow and applying dataflow analysis techniques to each run in order to infer all relevant data dependencies.

%% file: inn-sensitivity-full.tex
\newcommand{\RR}{\mathbb{R}}
\newcommand{\NN}{\mathbb{N}}
\newcommand{\ZZ}{\mathbb{Z}}
\newcommand{\QQ}{\mathbb{Q}}
\newcommand{\abs}[1]{\left|{#1}\right|}
\newcommand{\norm}[1]{|\!|{#1}|\!|}

The \emph{sensitivity of a function} is the expected maximum change in the output, given a change in the input of the function. Sensitivity is the basis for calibrating the amount of noise to be added to prevent leakages on statistical database queries using a differential privacy mechanism~\cite{Lee2011}. Differential privacy ensures that it is difficult for an attacker, who observes the query output, to distinguish between two input databases that are sufficiently ``close'' to each other, e.g. differ in one row. \pleak tells the user how to sample noise to achieve differential privacy, and how this affects the correctness of the output.

\pleak provides two methods -- global and local -- to quantify sensitivity of a task in a SQL workflow or of an entire SQL workflow. These methods can be applied to queries that output aggregations (e.g. count, sum, min, max).

\emph{Global sensitivity} analysis~\cite{globalsens} takes as input a database schema and a query, and computes the theoretical bounds for sensitivity, which are suitable for any instance of the database. Sensitivity shows how the output changes if we add (remove) a row to (from) some input table. To launch the analysis, the user clicks the ``Analyse Sensitivities'' button, receiving a matrix that shows the sensitivity w.r.t. each input table separately. 
It supports only COUNT queries.

Sometimes, the global sensitivity may be very large or even infinite. \emph{Local sensitivity} analysis is an alternative approach, which requires as input not only a schema and a query, but also a particular instance of the underlying database, and it tells how the output changes with the change \emph{from the given input}. Using the database instance improves the amount of noise needed to ensure differential privacy w.r.t. the number of rows. Moreover, it supports COUNT, SUM, MIN, MAX aggregations, and allows to capture more interesting distances between input tables, such as change in a particular attribute of some row. In \pleak, we have investigated a particular type of local sensitivity, called \emph{derivative sensitivity}~\cite{component}, which is in first place adapted to continuous functions, and is closely related to function derivative. \pleak uses derivative sensitivity to quantify the required amount of noise as described in~\cite{component}.

Let us look at some examples of derivative sensitivity analysis. Since differential privacy works with real-valued outputs, we cannot apply the analysis directly to the model of Fig.~\ref{fig:pe-bpmn-model}. We compute some related queries instead.

\begin{figure}[hbt]
    \includegraphics[width=1\textwidth]{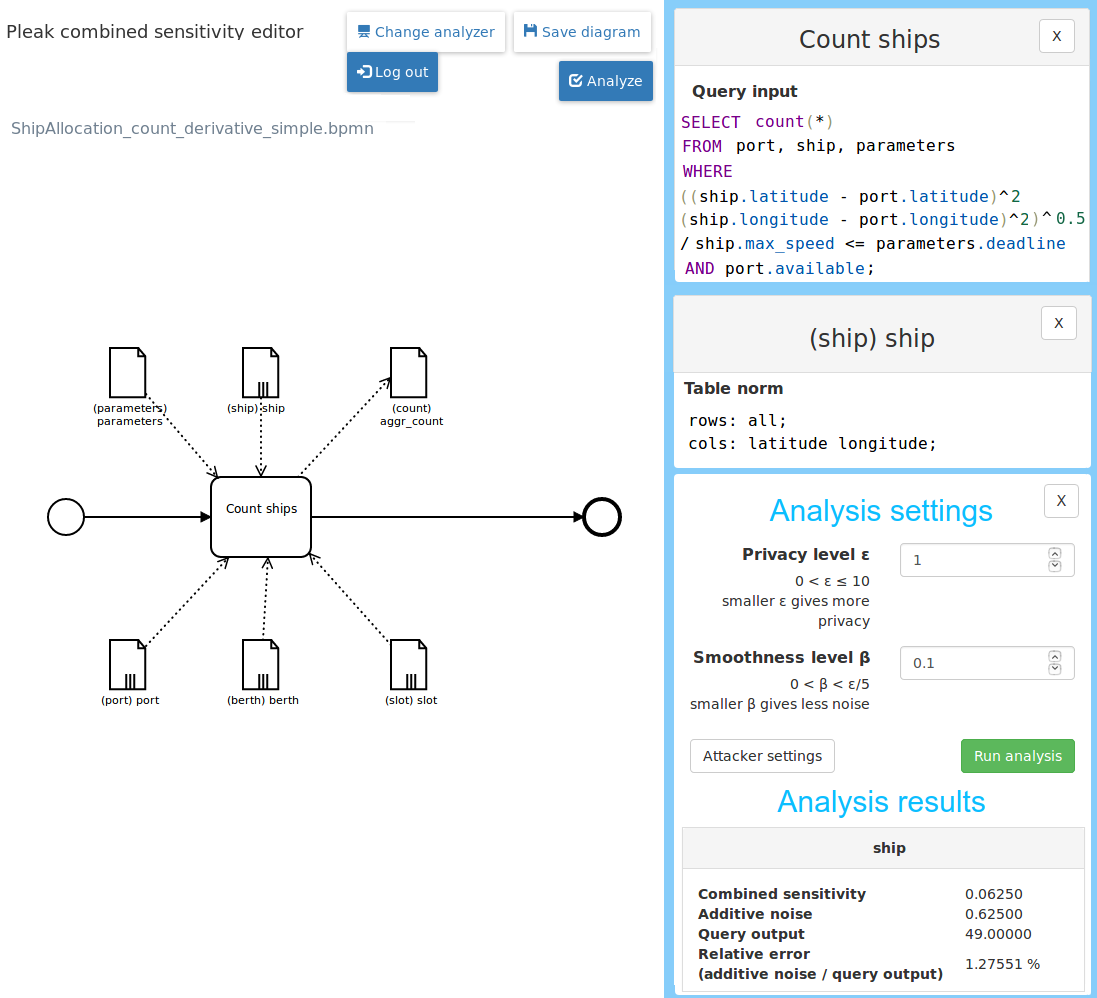}
    \caption{\label{fig:sensitivity}Derivative sensitivity analysis example: COUNT query}
\end{figure}

An example of derivative sensitivity analysis output with a COUNT query is shown in Fig.~\ref{fig:sensitivity}. The query counts the number of ships that are able to arrive at the available port before the deadline. The actual database instance contains 53 ships. The user wants to enforce differential privacy w.r.t. unit change in ship location (latitude and longitude), assuming that all ships (all rows in the table \emph{Ship}) are sensitive. This might correspond to the case where the user is the owner of the \emph{Ship} table, and the attacker is any other party that might see the output. The analysis result tells that the derivative sensitivity w.r.t. the \emph{Ship} table is $0.0625$, and that a differential privacy level of $\varepsilon=1$ can be achieved using smoothness parameter $\beta=0.1$. To this end, we would have to add an amount of noise such that the relative error of the output is $1.28\%$. More precisely, if the correct output is $y$, the noised answer will be between $0.9872 y$ and $1.0128 y$ with probability $80\%$.

\begin{figure}[hbt]
    \includegraphics[width=1\textwidth]{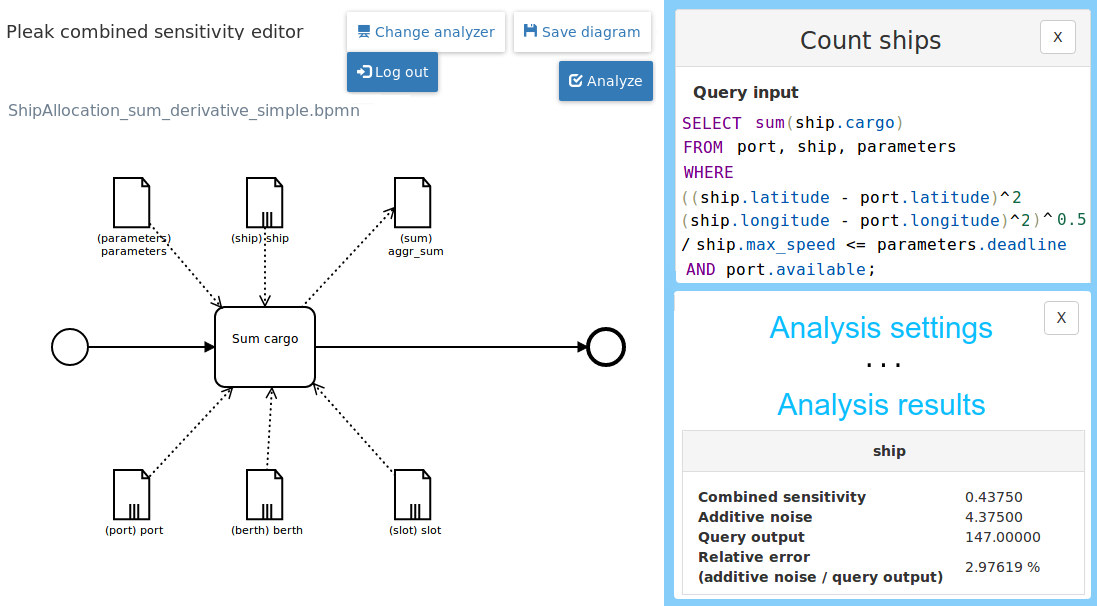}
    \caption{\label{fig:sensitivitySum}Derivative sensitivity analysis example: SUM query}
\end{figure}

A related SUM-query would be e.g. one that estimates the total amount of cargo that all arriving ships bring altogether. An example of a SUM query is shown in Fig.~\ref{fig:sensitivitySum}. The table norm and analysis settings are the same as in the COUNT query (Fig.~\ref{fig:sensitivity}) and are omitted from the figure. The sensitivity is larger, since some ships have more than $1$ unit of cargo and hence affect the output more, but the output itself is larger as well and in turn reduces the relative error.

\begin{figure}[hbt]
    \includegraphics[width=1\textwidth]{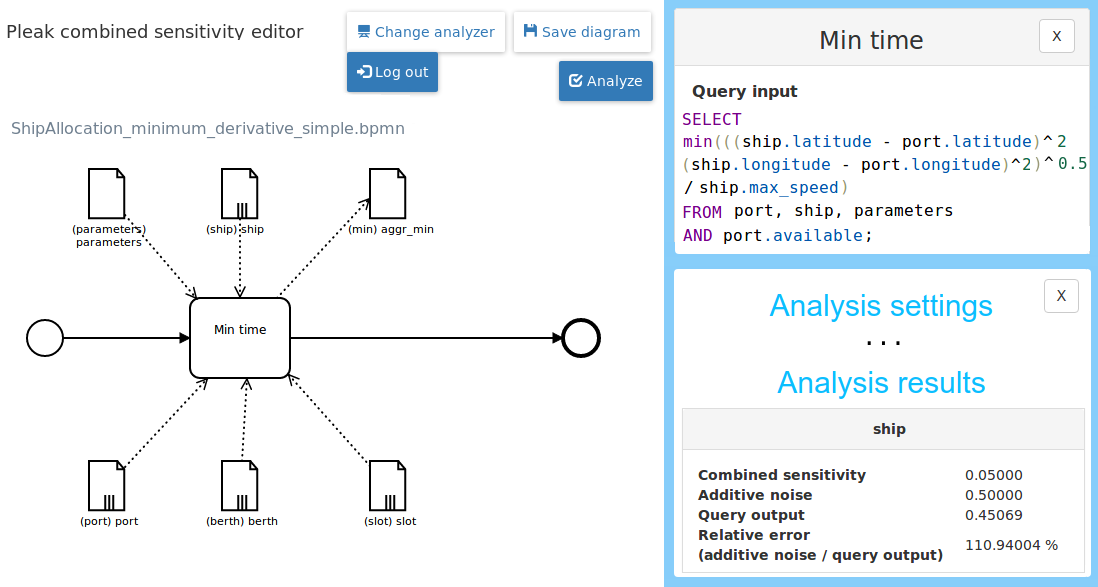}
    \caption{\label{fig:sensitivityMin}Derivative sensitivity analysis example: MIN query}
\end{figure}

Instead of counting the number of ships that reach the port before the deadline, we may be interested in the time when the first of them reaches the port. The corresponding example of a MIN query is shown in Fig.~\ref{fig:sensitivityMin}. The table norm and analysis settings are the same as before. We see that the error is quite large for a MIN query, and it is now $111\%$. While sensitivity itself is $0.05$, which is quite small, the reason why error is large is that the output itself is small. Differently from COUNT or SUM queries, the output does not increase with the number of table rows, and it is more difficult to achieve differential privacy.

\begin{figure}[hbt]
    \includegraphics[width=1\textwidth]{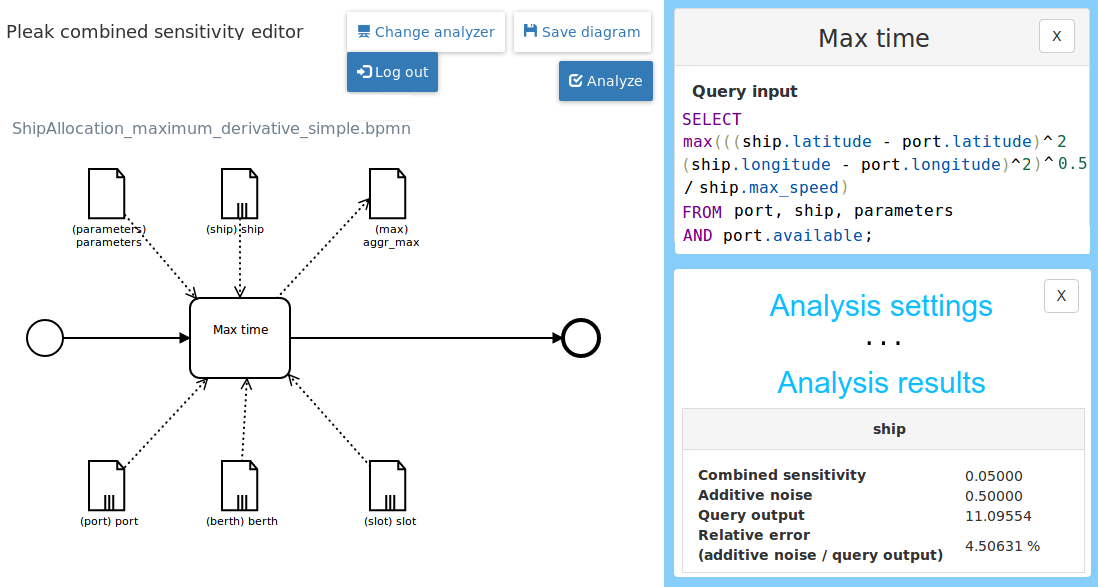}
    \caption{\label{fig:sensitivityMax}Derivative sensitivity analysis example: MAX query}
\end{figure}

It may be interesting to analyse a related query that computes the time when the \emph{last} ship reaches the port. The corresponding example of a MAX query is shown in Fig.~\ref{fig:sensitivityMax}. The table norm and analysis settings are the same as before. We see that the error is much smaller than for a MIN query, and it is $4.75\%$. This is because the output itself is large, so we in general would have smaller \emph{relative} errors for a MAX than for a MIN query over non-negative values (the absolute error remains the same).

A tutorial on sensitivity analyzer can be found at \url{https://pleak.io/wiki/sql-derivative-sensitivity-analyser}.

%% file: inn-advantage-full.tex
While function sensitivity as defined in Sec.~\ref{sec:sensitivity} can be used directly to compute the noise required to achieve $\varepsilon$-differential privacy, it is in general not clear which $\varepsilon$ is good enough, and the problem is that its ``goodness'' depends on the particular data and the query~\cite{Lee2011}. We want to use a more standard security measure, such as attacker's \emph{guessing advantage}. Formally, it is defined as the difference between the posterior (after observing the output) and prior (before observing the output) probabilities of attacker guessing the input. This tells the user how much the attacker is able to infer about the input after observing the output, in addition to what he has already known before (if anything). Internally, \pleak is still performing query function sensitivity analysis, but represents the analysis result in terms of guessing advantage, as described in~\cite{component}.

The \emph{guessing advantage} analysis of PLEAK takes as input the desired upper bound on attacker's advantage, which ranges between $0\%$ and $100\%$. The user specifies particular subset of attributes that the attacker is trying to guess for some data table record, within given precision range. To characterize the attacker more precisely, the user defines prior knowledge of the attacker, which is currently expressed as an upper and a lower bound on an attribute. The analyser internally converts these values to a suitable $\epsilon$ for differential privacy, and computes the noise required to achieve the bound on attacker's advantage.

\begin{figure}[ht]
    \includegraphics[width=1\textwidth]{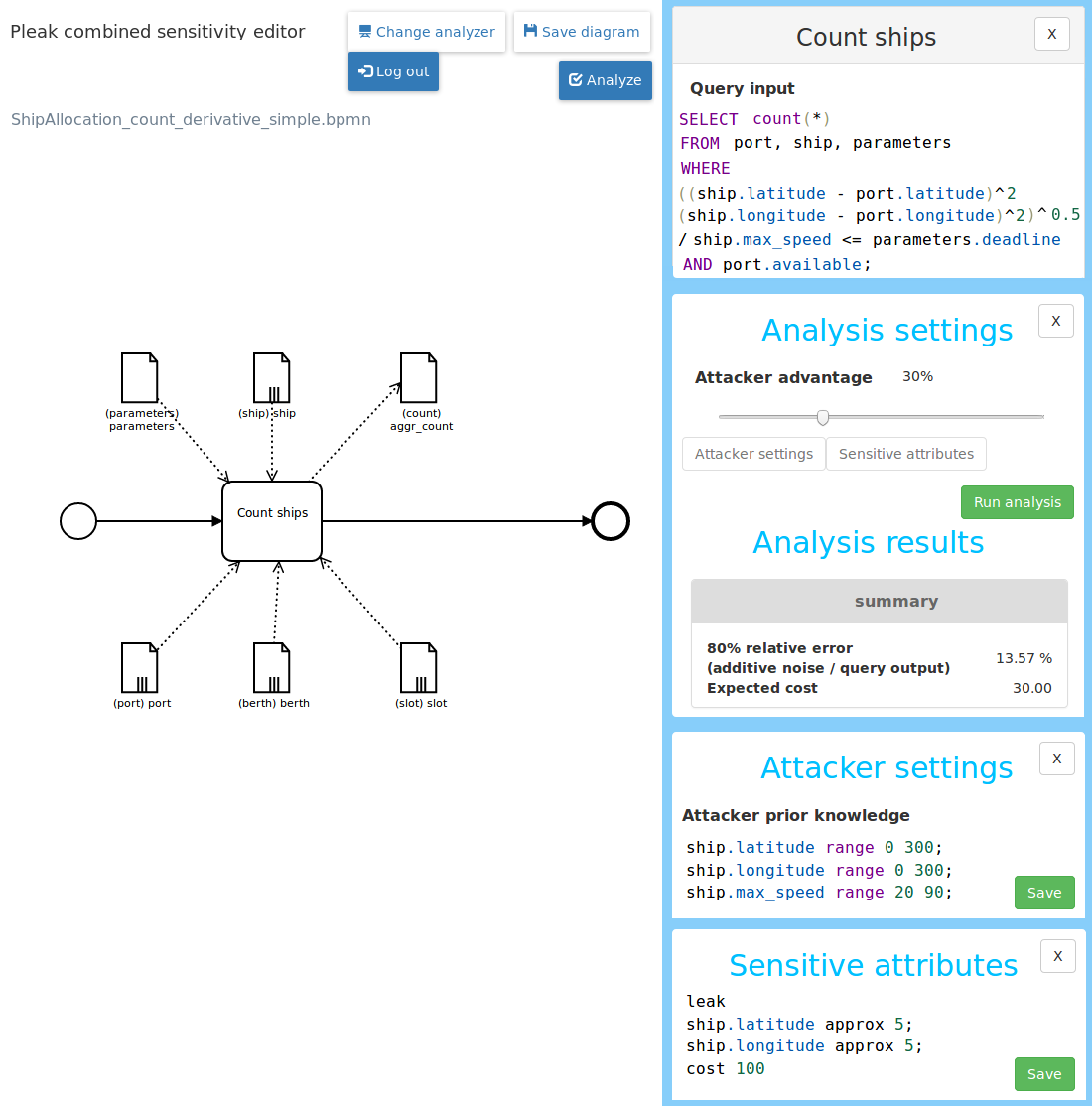}
    \caption{\label{fig:ga}Guessing advantage analysis example}
\end{figure}

Fig.~\ref{fig:ga} shows an example of guessing advantage analysis result. We consider the same COUNT query that has undergone sensitivity analysis in Fig.~\ref{fig:sensitivity}, which counts the total number of ships arriving before given deadline. Here, the attacker already knows that the longitude and latitude of a ship are in the range $[0..300]$ while the speed is in the range $[20..90]$. By default, he does not know anything else besides the bounds, and the prior distribution is assumed to be uniform in the range. Attacker's goal is to learn the location of any ship with a precision of $5$ units of its actual latitude and longitude. The analysis result says that, if we want to bound the guessing advantage by $30\%$ using noise addition mechanism, the relative error of the output will be $13.57\%$.

A tutorial on guessing advantage analyzer can be found at \url{https://pleak.io/wiki/sql-guessing-advantage-analyser}.
%, and this leakage costs $100$ cost units (e.g. dollars)
%, and the expected risk $30$ cost units

%% file: demo-plan.tex
PLEAK can be tried with username \emph{demo@example.com} and password \emph{pleakdemo}. The models created under this account are periodically deleted. In addition, PLEAK offers public view using the links included in the following. The public links are enough to see example models with their metadata and run PLEAK's analyzers. The account is necessary to create of modify the models.

The following description provides a walkthrough of capabilities of PLEAK using a unified scenario. The focus is on explaining the models and running the analysis and it is expected that the reader follows the writing using the demo account or the public links to the models. Our live demo would follow a similar pattern, but would allow for more interaction with the models, especially modifying the model data. Parts of the expected demonstration can be seen in the demo video in \url{https://www.youtube.com/watch?v=pQDYn1Q-BQM}. 

\subsection{Introduction to PLEAK}
The front page of pleak.io allows a user to log in and access its models using the files menu. Clicking on the model name in the file menu opens the editor used to create the BPMN model. Other actions can be accessed using the button in right hand side of the model row. Choosing the \emph{Shared models} tab also shows the models that are not owned by the user, but where others have granted either view or edit rights to the user. All models considered in this description are available for the demo account under the \emph{Shared models} tab with \emph{view} rights. The user can copy the shared models so that they appear in the \emph{My models} view and become modifiable. PLEAK also allows to publish models so that the analysis tools are accessible without a user account. 

PLEAK is built on top of the BPMN-js library and contains separate components to manage the model files, create BPMN models and several privacy-related editors. These editors can be accessed using the button on the right hand side of the model row in the file listing and they are the focus of this demo. 

All of the following revolves around a running scenario (e.g. see \url{https://pleak.io/app/#/view/Zta5dILQC6DozqcqQB4E}) that involves cargo ships and a nation with ports for the ships to dock at. The ship needs to find suitable berths available before its deadline. The data object \emph{reachable ports} contains ports that can be reached within the deadline. \emph{Feasible ports} narrows this down to ports that the ship can actually fit to. The final output of the process gives the actual port and berth slot assignment for each ship. The goal is to hide the ship location and the exact details of the ports where the ship can not dock.

The example models folder in the demonstration account has other processes that can be analyzed using our tools (model is intended to be used with the analyzer specified by the folder name). The process of using the tools is similar to the description given for the running example, but the concrete scenarios, the computations involved in the process, and therefore the analysis outputs can differ significantly. In addition, the wiki page in \url{pleak.io} gives further information about the usage and details of our tools.

\subsection{PE-BPMN}
Consider the Ship Allocation model using the PE-BPMN editor (\url{https://pleak.io/app/#/view/Zta5dILQC6DozqcqQB4E}). This is one possible process for agreeing on the slot assignment using secure multiparty computation (MPC). MPC methods allow participants to collaboratively compute on their data while only revealing the computation output. Privacy-Enhanced BPMN is a BPMN extension that captures the use of privacy enhancing technologies in the model. It adds notations to specify the technology and its concrete operation within a classical BPMN model, for example the blue \textsf{MPC} markers in the example.  

\begin{figure}
\centering
\includegraphics[width=0.8\textwidth]{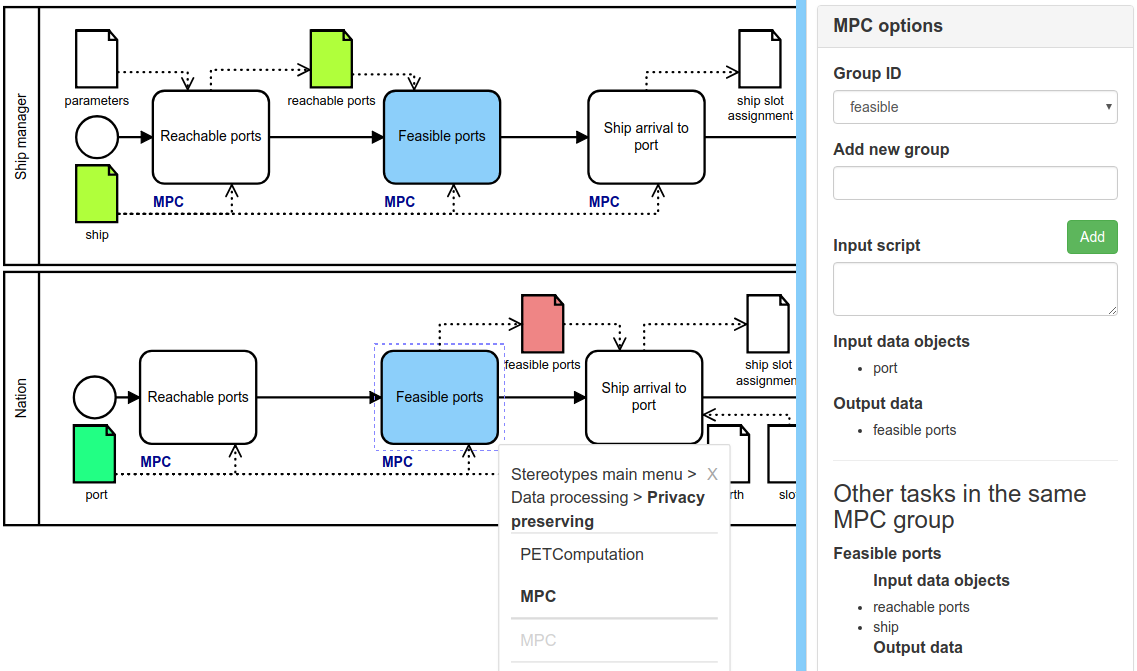}
\caption{Editing view of PE-BPMN editor}
\label{fig:pe-bpmn-editing}
\end{figure}

Clicking on tasks opens the stereotype menu when the user has \emph{edit} rights. This can be tried by copying the demo model so that it appears under \emph{My models} tab in the demo account and example of edit view of PE-BPMN editor is given in Fig.~\ref{fig:pe-bpmn-editing}. This menu is organized based on privacy goals such as data protection or processing. For example, MPC can be found under Data processing/Privacy preserving. Choosing a stereotype like secure multiparty computation opens a stereotype-specific panel on the right allowing to add required parameters. 

Tasks with the MPC stereotype are grouped based on which tasks correspond to a joint computation. The editor highlights the selected model element and other related elements, for example, other group members for MPC. In the given example, tasks with the same name in separate pools are considered to correspond to the same joint computation, hence, clicking on a task in one pool highlights the corresponding task in the other pool.

The correctness of PE-BPMN models can be checked using the \emph{Validate} button. The result of the validation appears on the right hand side of the screen. For a valid model, like the demo model, we get two analysis options - simple disclosure and data dependency. The simple disclosure report visualizes which participants have access to which data in the process. In addition, it distinguishes between data objects that are visible or hidden. For example, the Nation sees its inputs such as port, berth and slot and also learns the intermediate values, namely, feasible ports and the output assignment. However, the Nation does not have direct access to the inputs of the ship manager nor the reachable ports computed for the ship manager.
A data dependency matrix describes the interdependence of data objects. However, other analyzers offer more tools to go into the details of the dependencies. In the basic use of PLEAK, the analyst first finds potential leakages (data marked with \emph{V} in the disclosure report) and then uses the data dependency matrix to check if any visible data depends on any private data. If it does then the leaks-when or sensitivity analysis can be used to further study this dependency.

Validation produces an error list and does not allow analysis in case there are any problems in the model. For example, \url{https://pleak.io/app/#/view/NyWvwmKjUedE10nNyY6u} is an invalid model where clicking the validation button shows an error. Clicking on the error helps to locate the model elements that cause the error. In this case the second part of the \emph{feasible ports} task is missing the MPC stereotype so the error draws attention to the fact that the \emph{feasible ports} task in the \emph{Nation} requires another group member. The distributed nature of MPC tasks requires that there is at least two tasks in a group, hence the removal of one stereotype causes the remaining MPC task to give an error. 

Other demo account models consider different privacy enhancing technologies than MPC. Our approach to various PETs, including the concrete stereotypes and types of validation, is documented in PLEAK wiki\footnote{\url{https://pleak.io/wiki/pe-bpmn-editor_stereotypes}}.

\subsection{Leaks-When}
Open \url{https://pleak.io/app/#/view/lsQufWrKxjbdGtpJErHl} using the SQL editor to consider an example of the leaks-when analysis. The editor can be changed to SQL editor using the \emph{Change Analyzer} button in PLEAK.

Leaks-when analysis takes a SQL workflow or SQL collaborative workflow as an input. SQL workflow is a BPMN model where each task corresponds to SQL script that manipulates input database tables into temporary tables. The editor allows to view and edit these scripts. For example, clicking on the task \emph{Select reachable ports} reveals a SQL script that takes the tables \emph{port}, \emph{ship} and \emph{parameters} as inputs and produces \emph{reachable\_ports}. Data object are defined analogously, for the table \emph{port} one would enter an SQL CREATE TABLE statement. 

The data object \emph{parameters} is a special table that we use to define the name and data types of input parameters for the overall computation. In this scenario, we assume that the SQL-workflow is executed for one ship at a time such that the parameters are the ships name and desired deadline.

PLEAK's leaks-when analyzer processes the PostgreSQL's SQL dialect. For example, task \emph{Select feasible ports} has two store procedures, one of which computes the distance over the earth surface given the coordinates of two objects.

To analyze a fully annotated model, the analyst has to select one or more output data objects by clicking on them (selected data is green) and then start the analysis by clicking the button \emph{Leaks-when report}. For example, select \emph{feasible\_ports} and \emph{reachable\_ports} in the demo model. At the beginning of the analysis, PLEAK collects SQL scripts for each of the runs of the BPMN model and sends them to the backend for analysis.

The analysis output is a leaks-when report for each attribute. The right hand panel lists the chosen data objects and expanding the data object view shows the number of leakage graph corresponding to this data object. Each data object has one graph for each column in its output. For example, the \emph{reachable\_ports} data object has two columns, these correspond to the port and deadline computed in the script. The \emph{reachable\_ports(1)} is the port column. The final node in leaks-when graph is a filter where the first input shows what leaks and the second input shows under which conditions the leakage occurs. In \emph{reachable\_ports(1)} case, the leaks-when report shows that the \emph{port\_id} is disclosed if the ship can reach the port by a given deadline. The graph describes the deadline computation -- it is computed from ship's speed and distance from the port as determined from its coordinates. The second column \emph{reachable\_ports(0)} corresponds to deadline in the \emph{reachable\_ports} table. Looking at the corresponding leaks-when report shows that it leaks the arrival time under the same conditions as for the port. However, in the leaks branch of the graph we now also have the deadline computation.

The \emph{feasible\_ports} table is computed from the \emph{reachable\_ports} data and the leaks-when report reflects this. We can see that the deadline condition is still present for the leaks-when report of \emph{feasible\_ports}. In addition,  there are new conditions specific to this SQL query to stress that the ships draft has to be less than the harbor depth and that the port must be able to offload the cargo.

The leaks-when analysis can be extended to collaborative workflows. An example of our ships workflow as a collaboration can be seen in \url{https://pleak.io/app/#/view/wJuteo5sJAa_sf4cJ5oY}. 

\subsection{Sensitivity}
Sensitivity tells us how much information the output reveals about adding or modifying a row in the input table. Knowing the workflow's sensitivity allows us to make it differentially private (DP). PLEAK considers two flavours of sensitivity: global and local. Global sensitivity computes bounds based on the data structures whereas local sensitivity depends on the actual data.

\paragraph{Global.} Open \url{https://pleak.io/app/#/view/lh2NY01e2brJb6hspcFN}, Click the button \emph{Change Analyzer} and select \emph{SQL analyzer}. It is the same editor as used for leaks-when SQL analysis, and it requires similar SQL data object and task descriptions. Global sensitivity quantifies the magnitude of the noise that should be added to the output to make it differentially private. It can be computed based on the table schemas and the SQL workflow, similar to the leaks-when analysis.

Since global sensitivity is reasonable only for COUNT-queries, we count the number of ports for which the time that it takes to reach the port is below a certain threshold. Note that the main query does not contain the keyword COUNT, since the analyzer itself counts the rows in the output table.

The global sensitivity analysis starts by clicking the blue button \emph{Analyze Sensitivities}. The sensitivity matrix depicts the sensitivity of tasks in columns with respect to the input tables making up the rows. In this example, the sensitivity w.r.t. all tables except \emph{ship} is $\infty$. The sensitivity w.r.t. ship is $1$ since adding a ship may increase the total number of counted ships by $1$. If we remove the keyword DISTINCT from the query, the sensitivity becomes $\infty$, since now the same ship can be potentially assigned to an unbounded number of berths.

\paragraph{Local.} Open \url{https://pleak.io/app/#/view/xUILMd3SxrFF8wKPXV-7} in \emph{Combined sensitivity analyzer} to see both local and derivative sensitivities.

Similarly to SQL editor, this editor allows the user to define SQL statements for all tasks. In addition, it requires the user to insert actual input data tables to the data objects. Data can be viewed and added by clicking on the data objects. For example, select the \emph{ship} data object and consider the definition of sensitive rows in \emph{Table norm}. We see that all rows are considered sensitive (line \texttt{rows: all}), there are no sensitive columns (line \texttt{cols: none}), but the number of rows itself is sensitive, and the cost of adding/removing one row is 1.0 (line \texttt{G: 1.0}).

Analysis is started by clicking the button \emph{Analyze}. First, set the parameters to define the desired privacy level. The variable $\varepsilon$ comes directly from the definition of differential privacy. Simply put, a smaller epsilon means more privacy. The variable $\beta$ is a parameter that can be optimized. In general, it gives less noise if it is smaller but if it is too small, then achieving differential privacy may be impossible. The parameters can be left to their default values.

The analysis is executed by clicking on the green button \emph{Run analysis}. Similarly to global sensitivity, local sensitivity is computed with respect to each input table. In our case, the only sensitive table is \emph{ship}, and the sensitivity w.r.t. it is $2$. Indeed, since there is no keyword DISTINCT, a ship can be assigned to two possible berths of the port ``alma'' (assignments of berths to ports can be viewed by clicking the data object \emph{berth}), so the count may change by $2$. Recall that it would be $\infty$ in the case of global sensitivity. Relative error shows how much noise we have to tolerate to achieve differential privacy for \emph{ship} table.

%Let us try to change some parameters. Click the data object \emph{parameters} and write ``veis'' instead of ``alma'' in the table, this considering a different port. Click the green button \emph{Save}. If we now run the analysis, the sensitivity w.r.t. ship becomes $3$, since the port ``veis'' has $3$ berths.

\paragraph{Derivative.} Open \url{https://pleak.io/app/#/view/lQSSx15uY13H9S4EhcXA} in  \emph{Combined sensitivity analyzer} using \emph{Change Analyzer} button. This is the same model as before but the table norms are defined differently. In component based sensitivity, the user may choose which rows and columns are sensitive.

Click the \emph{port} data object and consider the definition of sensitive rows in \emph{Table norm}. Here we assume that the columns \emph{offloadcapacity}, \emph{offloadtime}, and \emph{harbordepth} are sensitive in all rows. It is possible to define more sophisticated sensitive components. In the table \emph{ship}, only the rows indexed $3$ and $7$ (ships ``gamma'' and ``farmi'') are considered sensitive as can be seen in the first row of the \emph{Table norm}. We combine latitude and longitude to define Euclidean distance (i.e $\ell_2$-norm) from the port with the line \texttt{u = lp 2.0 latitude longitude;}. We may assign different privacy weights to different columns, e.g. 0.2 in \texttt{v1 = scaleNorm 0.2 u;} means that we conceal changes in location up to $1 / 0.2 = 5$ units, so the location is more private than the length. Then, \texttt{z = lp 1.0 v1 v2;} means that the distance between two rows is the sum of distances between the location and the length (i.e $\ell_1$-norm). Finally, \texttt{return linf z;} shows how the distance between the tables is computed from the distances between their rows, and \texttt{linf} means that we take the maximum row distance (i.e $\ell_{\infty}$-norm), so DP conceals the change even if \emph{all} sensitive rows change by a unit. In general, DP requires much more noise to hide the changes in all rows simultaneously, but in our case only the rows $3$ and $7$ are sensitive, so it is fine.

Analysis is started by clicking the button \emph{Analyze}. The button \emph{Attacker settings} allows to define known bounds on table attributes. In the example model, it says that the ship maximum speed ranges from $20$ to $90$ units. Without the lower bound on ship speed, the arrival time approaches $\infty$ as speed approaches $0$, which does not allow to define a $\beta$-smooth lower bound for a finite $\beta$.

The analysis is executed by clicking on the green button \emph{Run analysis}. The sensitivity is computed with respect to each input table. The sensitivity w.r.t. table \emph{port} is very large. Indeed, if the port attributes change, it may happen that no ship will fit there anymore. Sensitivity w.r.t. the table ship is $4$, where $2$ comes from possible changes in the $3$rd row, and $2$ from possible changes in the $6$th row. The row sensitivity $2$ comes from the fact that modifying the length by $1$ unit, or the location by $5$ units, may cause filtering failure, and since there are $2$ available berths, we lose $2$ rows from the count.
 
\subsection{Guessing Advantage}

Open \url{https://pleak.io/app/#/view/P4RRkJV-DsBttt5NnapS}. Click the button \emph{Change Analyzer} and select \emph{Guessing advantage analyzer}. Here, each table has a schema and data, but no norm. Clicking \emph{Analyze} opens a slider, ranging from $0\%$ to $100\%$, to set the upper bound on attacker's guessing advantage. There are now two extra buttons to define bounds for used attributes:

\textbf{Attacker settings} defines prior knowledge of the attacker by setting pre-known bounds on attributes, defined either as \emph{exact}, \emph{range a b}, or \emph{total a} (the latter is used only for discrete data).

\textbf{Sensitive attributes} defines a set of sensitive components, which the attacker is trying to guess. The definition starts from a keyword \emph{leak} and for each attribute, the guess can either be \emph{exact} (discrete attributes), or \emph{approx r} (approximated by $r > 0$ units). The list of attributes is followed by the keyword \emph{cost} and a number that defines the cost of leaking that attribute.

%By default, the advantage slider is set to $30\%$. 
Reducing the advantage slider to $0\%$ gives the error $\infty$, as it is impossible to achieve perfect privacy with bounded noise. Increasing it to $100\%$ gives the error $0$, since the attacker is allowed to guess everything. Reducing the allowed guessing radius under \emph{Sensitive attributes}, or the known radius under \emph{Attacker settings} (click \emph{Save} after making any changes) makes the guess more difficult. Clicking \emph{View more}, we see that both prior and posterior probabilities decrease. However, since the noise level depends on the \emph{advantage}, which is the difference between these two probabilities, the error does not necessarily decrease.